\documentclass[3p,times,sort&compress]{elsarticle}

\usepackage[colorlinks = true,
            linkcolor = blue,
            urlcolor  = blue,
            citecolor = blue,
            anchorcolor = blue,
            unicode]{hyperref}

\usepackage{graphicx}
\usepackage{amsmath}
\usepackage{amssymb}
\usepackage{amsfonts}
\usepackage{color}

\renewcommand{\Re}{\mathop\mathrm{Re}\nolimits}
\renewcommand{\Im}{\mathop\mathrm{Im}\nolimits}

 \DeclareMathOperator{\arctanh}{arctanh}

\begin{document}

\begin{frontmatter}

\title{Peculiarities of the density of states in SN junctions}

\author[MIPT,JINR]{A.~A.\ Mazanik}

\author[ITP,HSE]{Ya.~V.\ Fominov\texorpdfstring{\corref{cor1}}{}}
\ead{fominov@itp.ac.ru}

\address[MIPT]{Moscow Institute of Physics and Technology, 141700 Dolgoprudny, Russia}
\address[JINR]{BLTP, Joint Institute for Nuclear Research, 141980 Dubna, Russia}
\address[ITP]{L.~D.\ Landau Institute for Theoretical Physics RAS, 142432 Chernogolovka, Russia}
\address[HSE]{Laboratory for Condensed Matter Physics, HSE University, 101000 Moscow, Russia}

\cortext[cor1]{Corresponding author}

\begin{abstract}
We study the density of states (DoS) $\nu(E)$ in a normal-metallic (N) film contacted by a bulk superconductor (S). We assume that the system is diffusive and the SN interface is transparent.
In the limit of thin N layer (compared to the coherence length), we analytically find three different types of the DoS peculiarity at energy equal to the bulk superconducting order parameter $\Delta_0$.
(i)~In the absence of the inverse proximity effect,  the peculiarity has the check-mark form with $\nu(\Delta_0)=0$ as long as the thickness of the N layer is smaller than a critical value.
(ii)~When the inverse proximity effect comes into play, the check-mark is immediately elevated so that $\nu(\Delta_0)>0$.
(iii)~Upon further increasing of the inverse proximity effect, $\nu(E)$ gradually evolves to the vertical peculiarity
(with an infinite-derivative inflection point at $E=\Delta_0$).
This crossover is controlled by a materials-matching parameter which depends on the relative degree of disorder in the S and N materials.
\end{abstract}

\date{27 December 2022}


\end{frontmatter}

\tableofcontents

\section{Introduction}
\label{sec:Intro}

\begin{figure}[t]
 \centerline{\includegraphics[width=0.6\columnwidth]{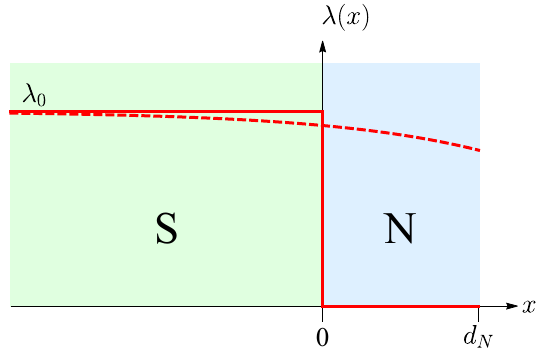}}
\caption{Superconductor/normal metal (SN) junction. The superconductor occupies a half-space, the N layer has thickness $d_N$. The BCS pairing constant $\lambda(x)$ is a step function (vanishing in the N layer) shown by the solid red line. For comparison, we also show schematically a $\lambda(x)$ dependence corresponding to the model of Ref.\ \cite{Fominov2019.PhysRevB.100.224513}: a superconductor with weak surface suppression of the pairing constant (dashed red line).}
 \label{fig:SN}
\end{figure}

When a normal-metallic (N) film is deposited on the surface of a superconductor (S), see Fig.~\ref{fig:SN}, it acquires some superconducting properties. This is the essence of the superconducting proximity effect \cite{deGennes1964.RevModPhys.36.225,McMillan1968,Deutscher1969}. The superconducting correlations induced into the N layer modify, in particular, the quasiparticle density of states (DoS) $\nu(E)$ (measured in units of the normal-state DoS).
The DoS in various SN systems can be theoretically calculated \cite{Golubov1988,Golubov1989,Golubov1996,Belzig1996.PhysRevB.54.9443,Zhou1998,Wilhelm2000.PhysRevB.62.5353,Altland2000,Cuevas2006.PhysRevB.73.184505,
Hammer2007.PhysRevB.76.064514,Gurevich2017} and
experimentally measured
\cite{Gueron1996.PhysRevLett.77.3025,Pannetier2000,Scheer2001.PhysRevLett.86.284,Vinet2001.PhysRevB.63.165420,Moussy2001,leSueur2008.PhysRevLett.100.197002,
Meschke2011.PhysRevB.84.214514}
with the help of the point-contact tunneling spectroscopy or the scanning tunneling spectroscopy.
In the diffusive limit, the proximity-induced DoS in the N layer is characterized by an energy gap $E_g$ (with $E_g<\Delta_0$).

Another side of the proximity effect is partial suppression of superconductivity on the S side. Both effects (direct proximity effect in the N part and inverse proximity effect in the S part) take place mainly in the vicinity of the interface. Their characteristic length scales are the coherence lengths \cite{Belzig1996.PhysRevB.54.9443}
\begin{equation}
\xi_N = \sqrt{D_N/2\Delta_0},\qquad \xi_S = \sqrt{D_S/2\Delta_0},
\end{equation}
where $D_{N(S)}$ is the diffusion constant in the N (or S) layer and $\Delta_0$ is  the bulk superconducting gap.
The proximity effect in the N part is strongest in the limit of transparent SN interface and thin N layer, $d_N\ll \xi_N$; the induced energy gap $E_g$ is then close to $\Delta_0$ (we will consider systems with half-infinite S part so that superconductivity is not perturbed in its bulk).

In addition to transparency, there is one more important interface parameter which describes materials matching,
\begin{equation} \label{eq:varkappa}
\varkappa = D_S \sigma_N^2 / D_N \sigma_S^2
\end{equation}
(in the case when the normal-state DoS of the S and N material coincide, we would end up with $\varkappa=\sigma_N/\sigma_S$).
It determines ``softness'' of the superconductor: at large $\varkappa$, the S part is soft in the sense that superconductivity is essentially suppressed in the SN interface region (strong inverse proximity effect), while at small $\varkappa$, the S part is rigid in the sense that bulk superconducting characteristics are only slightly altered in the vicinity of the interface (weak inverse proximity effect).

Note that while the DoS in the N layer is a local (spatially-dependent) quantity, the energy gap characterizes the N layer as a whole. When discussing the energy dependence of the DoS, for definiteness we will consider the surface DoS (taken at the outer surface and thus directly available for, e.g., STM measurements). At the same time, qualitative features of the $\nu(E)$ dependence persist at any point inside the N layer; this is especially clear in the limit of thin N layer.

\begin{figure}[t]
     \centerline{\includegraphics[width=0.9\columnwidth]{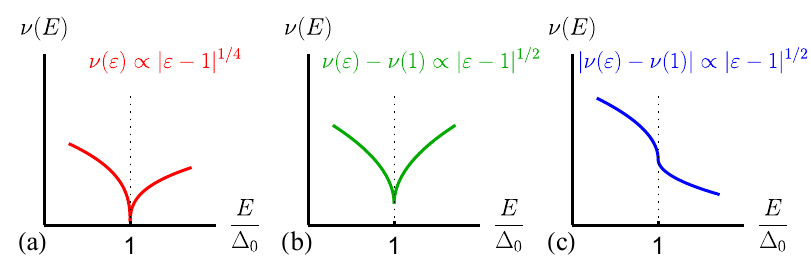}}
  \caption{Schematic plot of three different types of the DoS peculiarity at $E=\Delta_0$: (a)~full check-mark peculiarity, (b)~elevated check-mark peculiarity, (c)~vertical peculiarity.
  As discussed later in Section~\ref{sec:SNwithinvprox}, the elevated check-mark peculiarity is symmetric, while the full check-mark and the vertical peculiarity are asymmetric.}
 \label{fig:peculiarities}
\end{figure}

In addition to the presence of the gap $E_g$, numerically calculated DoS in the N layer usually demonstrates a peculiarity at $E=\Delta_0$ (the energy scale inherited from the superconductor), the form of which depends on parameters of the system
\cite{Golubov1988,Golubov1989,Golubov1996,Belzig1996.PhysRevB.54.9443,Wilhelm2000.PhysRevB.62.5353,Altland2000,Hammer2007.PhysRevB.76.064514,Gurevich2017}.
Some time ago, a striking form of peculiarity was analytically derived by Levchenko \cite{Levchenko2008}: he predicted that in the limit of thin N layer and transparent interface, the DoS in the N layer demonstrates a ``check-mark'' behavior around $E=\Delta_0$ with $\nu(E) \propto |E-\Delta_0|^{1/4}$, see Fig.~\ref{fig:peculiarities}(a). This type of behavior is rather unexpected since it implies vanishing of the DoS not only below $E_g$ but also exactly at $E=\Delta_0$.

Recently, we studied peculiarity of the DoS at $E=\Delta_0$ in a different system, a superconductor with surface suppression of the BCS pairing constant $\lambda(x)$ \cite{Fominov2019.PhysRevB.100.224513}, see Fig.~\ref{fig:SN}. In the limit of \emph{weak} suppression of $\lambda(x)$ on a short spatial scale $r_c \ll \xi_S$ near the surface, we analytically found the ``vertical'' peculiarity of the DoS [monotonically decreasing $\nu(E)$ with an infinite-derivative inflection point at $E=\Delta_0$ and large value $\nu(\Delta_0)\gg 1$, see Fig.~\ref{fig:peculiarities}(c)]. This is completely different from the check-mark peculiarity but at first sight this is not surprising since the two types of peculiarity arise in different systems. However, it turns out that comparison is possible. An SN system with $d_N \ll \xi_N$ can be considered as a system with short-scale surface suppression of $\lambda(x)$ (which is a step function equal to the bulk value $\lambda_0$ in the S part and to $0$ in the N part). At the same time, this implies \emph{strong} surface suppression of $\lambda(x)$ (strong deviation from the bulk value). On the other hand, the assumption of weak suppression in Ref.\ \cite{Fominov2019.PhysRevB.100.224513} was needed only for analytical implementation of self-consistency for the order parameter $\Delta(x)$. If the self-consistency is neglected (as in Ref.\ \cite{Levchenko2008}), then the analytical method of Ref.\ \cite{Fominov2019.PhysRevB.100.224513}  is applicable for SN structures as well ($r_c$ in this case should be identified with $d_N$, see Fig.~\ref{fig:SN}). Therefore, the results of Ref.\ \cite{Fominov2019.PhysRevB.100.224513} predict the vertical peculiarity of the DoS in SN system with equivalent S and N materials (which differ only by the pairing constant $\lambda$; this case corresponds to $\varkappa=1$).

We are therefore faced with contradicting predictions about the form of the DoS peculiarity in the SN system. The necessity to resolve the contradiction is the motivation for the present work.
Note that in addition to the fact that the surface DoS can be directly probed experimentally \cite{Gueron1996.PhysRevLett.77.3025,Pannetier2000,Scheer2001.PhysRevLett.86.284,Vinet2001.PhysRevB.63.165420,Moussy2001,leSueur2008.PhysRevLett.100.197002},
it also directly influences various physical properties, e.g., the tunneling current \cite{Meschke2011.PhysRevB.84.214514} and the surface impedance \cite{TinkhamBook,Gurevich2017,Kubo2019}.
Some experiments demonstrate the DoS suppression in SN junctions at $E=\Delta_0$ by means of scanning tunneling microscopy \cite{leSueur2008.PhysRevLett.100.197002} or tunneling spectroscopy \cite{Meschke2011.PhysRevB.84.214514}.
In superconducting qubit devices, local subgap DoS (at $E<\Delta_0$) can work as quasiparticle traps, mitigating the adverse effect of quasiparticles on the coherence \cite{Riwar2019.PhysRevB.100.144514}.

Below, we reconsider the results of Ref.\ \cite{Levchenko2008} and demonstrate that the full check-mark behavior \cite{Levchenko2008} of the surface DoS [with $\nu(\Delta_0)=0$, see Fig.~\ref{fig:peculiarities}(a)] is indeed realized in SN junctions with thin N layer but only in the limit of absolutely rigid superconductor, $\varkappa = 0$. At the same time, $\nu(\Delta_0)$ becomes finite at any finite $\varkappa$. The check-mark peculiarity is elevated at $\varkappa\neq 0$ [see Fig.~\ref{fig:peculiarities}(b)] and finally crosses over to the vertical peculiarity at $\varkappa \gtrsim (d_N/\xi_N)^4$ [see Fig.~\ref{fig:peculiarities}(c)]. Since $d_N/\xi_N\ll 1$, the vertical peculiarity is realized at $\varkappa=1$ \cite{Fominov2019.PhysRevB.100.224513}.

We therefore clarify the existing contradiction between previously reported results and describe continuous evolution between qualitatively different types of the DoS peculiarity with varying the softness parameter $\varkappa$.
The elevated check-mark behavior turns out to be the ``missing element'' providing this crossover.
This type of peculiarity has not been analytically described before, to the best of our knowledge.

Aiming at resolving the existing contradiction, we pay special attention to comparison with previous publications wherever possible in order to underline not only new results but also agreements with previously reported results (as well as their corrections and generalizations).

The paper is organized as follows:
In Section~\ref{sec:Model}, we formulate equations of the quasiclassical theory in the diffusive limit, which are relevant to our model.
In Section~\ref{sec:SNwithinvprox}, we apply the general equations to analytically calculate the surface DoS in thin N layer in the limiting cases of rigid and soft superconductor.
In Section~\ref{sec:FiniteThickness}, we analyze modification of the DoS in the limit of absolutely rigid superconductor as the N layer thickness grows.
In Section~\ref{sec:numerics}, we illustrate and generalize our analytical calculations by numerical results.
In Section~\ref{sec:Discussion}, we discuss possibility of experimental observation of the predicted DoS peculiarities.
In Section~\ref{sec:Conclusions}, we present our conclusions.
Finally, some details of calculations are presented in the Appendixes.

Throughout the paper, we employ the units with $\hbar = 1$.

\section{Model}
\label{sec:Model}

The derivation of this section reproduces the derivation of Ref.\ \cite{Levchenko2008}. We present it in order to establish notations and underline important points that will be essential for future analysis and explanation of difference between our results and results of Ref.\ \cite{Levchenko2008}.

\subsection{General equations}

We consider a diffusive system shown in Fig.~\ref{fig:SN}, which is a normal metallic layer (at $0 < x < d_N$) contacted by an $s$-wave superconductor (at $x < 0$).
To calculate the DoS in this inhomogeneous system, we employ the quasiclassical approach \cite{Usadel1970,Belzig1999}.
The system and the theoretical approach are the same as in Ref.\ \cite{Levchenko2008} (our SN system is a half of the SNS junction of Ref.\ \cite{Levchenko2008}, which does not influence the solution).

The quasiclassical method in the theory of superconductivity is based on smallness of the superconducting energy scale $\Delta_0$ compared to the Fermi energy (or, equivalently, smallness of the Fermi wavelength compared to the superconducting coherence length) \cite{Usadel1970,Rammer1986.RevModPhys.58.323,LarkinOvchinnikov1986NoneqScReview,Belzig1999,KopninBook}. The Gor'kov equations describe superconductivity in the language of the Green functions with the help of the conventional (normal) $G$ function (describing electrons) and the Gor’kov (anomalous) $F$ function (describing Cooper pairs) \cite{AGDBookEng}. Within the framework of the quasiclassical approach, the equations can be simplified (physically, this implies averaging over atomic-scale oscillations), and take the form of the Eilenberger--Larkin--Ovchinnikov equations \cite{Eilenberger1968,Larkin1968}. These equations can include effects of impurity scattering, and in the diffusive limit (mean free path much smaller than the superconducting coherence length) turn into the Usadel equation.

With the help of the standard $\theta$ parametrization \cite{Stoof1996.PhysRevB.53.14496,Belzig1999}, we can write the normal and anomalous Green functions of the quasiclassical theory as $G=\cos\theta$ and $F=\sin\theta$, respectively. The Usadel equation in the two parts of the system then takes the form
\begin{gather}
\frac{D_S}2 \frac{d^2 \theta_S}{dx^2} - \sqrt{\Delta_0^2-E^2} \sin(\theta_S-\theta_\mathrm{BCS}) =0,
\label{eq:UsadelS} \\
\frac{D_N}2 \frac{d^2 \theta_N}{dx^2} +i E \sin\theta_N =0,
\end{gather}
where the bulk (BCS) solution in the S part is
\begin{equation}
\theta_\mathrm{BCS} = \pi/ 2 +i\arctanh ( E/\Delta_0).
\end{equation}

The boundary conditions at the transparent SN interface ($x=0$) ensures continuity of the Green functions and of the current,
while the boundary condition at the outer surface of the N layer ($x=d_N$) ensures absence of the current \cite{Kupriyanov1988}:
\begin{equation}
\theta_S(0) = \theta_N(0),\qquad \sigma_S \frac{d \theta_S(0)}{dx} = \sigma_N \frac{d\theta_N(0)}{dx},
\qquad
\frac{d\theta_N(d_N)}{dx} =0.
\label{eq:bc_x=dN}
\end{equation}

In order to demonstrate the role of the materials-matching parameter $\varkappa$ [defined by Eq.\ \eqref{eq:varkappa}], we can normalize the coordinate in each part of the structure by the corresponding coherence length, and then rewrite the second of the interface boundary conditions in Eq.\ \eqref{eq:bc_x=dN} as
\begin{equation}
\frac{d\theta_S(0)}{d(x/\xi_S)} = \sqrt{\varkappa} \frac{d \theta_N(0)}{d(x/\xi_N)}.
\end{equation}
The limit of $\varkappa\to 0$ corresponds to absolutely rigid superconductor (no inverse proximity effect), while the limit of $\varkappa\to \infty$ corresponds to soft  superconductor (strong inverse proximity effect).

Note that in Eq.\ \eqref{eq:UsadelS}, we have neglected self-consistency, assuming the order parameter in the form of the step function $\Delta(x) = \Delta_0$ in the S part while $0$ in the N part. Actually, $\Delta(x)$ can be suppressed in the vicinity of the interface on the S side. This is done in order to underline comparison with the results of Ref.\ \cite{Levchenko2008} (where self-consistency was neglected). As we see below, the check-mark behavior predicted in Ref.\ \cite{Levchenko2008} actually takes place only in the limit $\varkappa\to 0$, where neglecting self-consistency is fully justified. On the other hand, at finite $\varkappa$ self-consistency does not lead to qualitative changes, in particular, it does not eliminate peculiarity at $E=\Delta_0$ (as we can see from comparison with results of Ref.\ \cite{Fominov2019.PhysRevB.100.224513} which correspond to $\varkappa=1$ and where self-consistency was rigorously taken into account).

At the same time, the set of equations \eqref{eq:UsadelS}--\eqref{eq:bc_x=dN} generally describe both the direct ($\theta_N\neq 0$) and inverse ($\theta_S \neq \theta_\mathrm{BCS}$) proximity effect in terms of the Green functions (and all physical properties, such as the DoS, following thereof).

We will be interested in the DoS at the outer surface of the N layer:
\begin{equation} \label{eq:DoSdef}
    \nu_N(E) \equiv \left. \nu(E) \right|_{x=d_N} = \Re \cos\theta_N(d_N).
\end{equation}

\subsection{Effective equations for the N layer}
\label{sec:Neff}

It is convenient to rewrite the equations in real form with the help of transformation
\begin{equation}
 \theta_N = \pi/2 + i\psi_N,
 \qquad
 \theta_S = \theta_\mathrm{BCS}+i\psi_S.
\end{equation}
Equations  \eqref{eq:UsadelS}--\eqref{eq:bc_x=dN} then take the form
\begin{gather}
\frac{D_S}2 \frac{d^2 \psi_S}{dx^2} -\sqrt{\Delta_0^2-E^2} \sinh\psi_S =0,  \\
\frac{D_N}2 \frac{d^2 \psi_N}{dx^2} +E \cosh\psi_N =0, \\
\psi_N(0) = \psi_S(0)+\arctanh (E/\Delta_0), \\
\sigma_N \frac{d\psi_N(0)}{dx} = \sigma_S \frac{d\psi_S(0)}{dx},
\qquad
\frac{d\psi_N(d_N)}{dx} =0.
\end{gather}

The solution in the S part decaying at $x\to -\infty$ is known and can be parametrized by a single parameter, the interface value $\psi_S(0)$:
\begin{equation}
  \psi_S(x) = 4 \arctanh\left[ \tanh\left( \psi_S(0)/4\right) \exp\left( x/\xi_E \right) \right],
\end{equation}
where
\begin{equation} \label{eq:xiE}
  \xi_E = \frac{\xi_S}{\left[ 1-(E/\Delta_0)^2\right]^{1/4}}.
\end{equation}

We define dimensionless coordinate (in the N layer) and energy,
\begin{equation}
X= x/d_N,
\qquad
\varepsilon = E /\Delta_0.
\end{equation}
The thickness of the N layer determines the Thouless energy which can then be used to define dimensionless order parameter:
\begin{equation} \label{eq:ETh_delta0}
E_\mathrm{Th} = D_N/d_N^2,
\qquad
\delta_0 = \Delta_0/E_\mathrm{Th} = (d_N/\xi_N)^2 /2.
\end{equation}

The set of equations for $\psi_N(X,\varepsilon)$ then takes the form
\begin{gather}
\psi_N''/2 +\varepsilon \delta_0 \cosh\psi_N = 0, \label{eq:psiNUs}
\\
\psi_N(0) = \psi_S(0) + \arctanh\varepsilon, \label{eq:psi_N(0)}
\\
\psi_N'(0) = \sqrt{8\delta_0/\varkappa} (1-\varepsilon^2)^{1/4} \sinh\left[ \psi_S(0)/2 \right], \label{eq:psi_N'(0)}
\\
\psi_N'(1) = 0. \label{eq:psiNbc}
\end{gather}
The derivatives here are taken with respect to $X$ and we omit $\varepsilon$ in the argument of $\psi_N$ for brevity.

The obtained equations are for the N layer only. All information about the solution in the S part enters only through the interface value $\psi_S(0)$ which can be easily eliminated from Eqs.\ \eqref{eq:psi_N(0)} and \eqref{eq:psi_N'(0)}.

The DoS \eqref{eq:DoSdef} at the outer surface of the N layer is now given by
\begin{equation} \label{eq:nu_N}
    \nu_N(\varepsilon) = \Im \sinh\psi_N(1).
\end{equation}

\subsection{Equations in the limit of thin N layer}
\label{sec:thinN}

The limit of thin N layer is defined by condition $d_N\ll \xi_N$, or, equivalently,
\begin{equation} \label{eq:delta0<<1}
\delta_0 \ll 1.
\end{equation}
In this limit, the proximity-induced gap $E_g$ should be only slightly smaller than $\Delta_0$ (in dimensionless units, $\varepsilon_g \equiv E_g/\Delta_0 \approx 1$).

The hyperbolic arctangent in the right-hand side (r.h.s.) of Eq.\ \eqref{eq:psi_N(0)} tends to infinity at $\varepsilon\to 1$. One can therefore expect \cite{Levchenko2008} that
\begin{equation} \label{eq:approx}
\Re \psi_N \gg 1,
\qquad
\cosh\psi_N \approx e^{\psi_N}/2
\end{equation}
at the interface ($x=0$) and, moreover, that these relations are satisfied everywhere inside the N layer (due to its small thickness).
We expect that this approximation is suitable not only for analysis of the DoS at $\varepsilon=1$ but also for calculating the gap $\varepsilon_g$.
Validity of this approximation must be checked after the calculations are performed.

We can now simplify Eq.\ \eqref{eq:psiNUs} as
\begin{equation} \label{eq:Ussimplified5}
\psi_N'' +\varepsilon \delta_0 e^{\psi_N} = 0,
\end{equation}
which can be solved in terms of elementary functions \cite{Levchenko2008}:
\begin{equation} \label{eq:psiN5}
\psi_N(X) = \psi_N(1) - \ln \cosh^2\left[ e^{\psi_N(1)/2} \sqrt{\varepsilon\delta_0 /2} (X-1) \right].
\end{equation}
The boundary condition at $X=1$, Eq.\ \eqref{eq:psiNbc}, has already been taken into account here.
The remaining boundary conditions at $X=0$, Eqs.\ \eqref{eq:psi_N(0)} and \eqref{eq:psi_N'(0)}, yield two equations for two parameters, $\psi_N(1)$ and $\psi_S(0)$. Excluding $\psi_S(0)$, we obtain a single equation for $\psi_N(1)$.
In terms of a new variable
\begin{equation}
    V =\sqrt{\varepsilon\delta_0 /2} \exp\left( \psi_N(1)/2 \right),
\end{equation}
this equation can be written as
\begin{equation}
  \sqrt{\varkappa} \sinh V +\sqrt{\varepsilon (1+\varepsilon)} \frac{\delta_0 \cosh^2 V}{2 V^2}= \sqrt{\frac{1-\varepsilon}{\varepsilon}}.
\end{equation}
Assuming
\begin{equation} \label{eq:epscloseto1}
|1-\varepsilon|\ll 1,
\end{equation}
we finally simplify the equation as
\begin{equation} \label{eq:V}
  \sqrt{\varkappa} \sinh (V)  + \frac{\delta_0 \cosh^2 V}{\sqrt{2} V^2} = \sqrt{1-\varepsilon}.
\end{equation}
Our assumption that $\Re\psi_N \gg 1$ everywhere inside the N layer implies
\begin{equation} \label{eq:Vcond}
|V|\gg \sqrt{\delta_0}.
\end{equation}

Within our current accuracy, the DoS \eqref{eq:nu_N} at the outer surface of the N layer is given by
\begin{equation} \label{eq:DoSgen}
  \nu_N(\varepsilon) =
  \Im V^2 / \delta_0. 
\end{equation}

The approximations we have used up to now are given by Eqs.\ \eqref{eq:approx} [its consequence in terms of $V$ is given by Eq.\ \eqref{eq:Vcond}] and \eqref{eq:epscloseto1}.

The softness parameter $\varkappa$ determines two limiting cases of ``rigid'' and ``soft'' superconductor (strong and weak superconductor, in terminology of Levchenko \cite{Levchenko2008}). As we will see below, they correspond to conditions $\varkappa \ll \delta_0^2$ and $\varkappa\gg \delta_0^2$, respectively.

\section{DoS in thin N layer}
\label{sec:SNwithinvprox}

In this section, we solve Eq.\ \eqref{eq:V} and calculate the DoS given by Eq.\ \eqref{eq:DoSgen} in the limiting cases of rigid and soft superconductor. In each limiting case, we analyze peculiarity of the DoS at $E\to\Delta_0$ and find the energy gap $E_g$ as well as the behavior of the DoS at $E\to E_g$.

In each case, we check validity of condition \eqref{eq:Vcond}. Other necessary checks are discussed in 
\ref{app:checks}.

\subsection{Limit of rigid superconductor, \texorpdfstring{$\varkappa \ll \delta_0^2$}{kappa<<delta0squared}}
\label{sec:rigid}

\subsubsection{\texorpdfstring{$E\to\Delta_0$}{E->Delta0}}
\label{sec:rigidEtoDelta}

From Eq.\ \eqref{eq:V}, we see that at $\varkappa\to 0$ and $\varepsilon\to 1$, we have $V\to i\pi/2$ [which implies that condition \eqref{eq:Vcond} is satisfied as it should be for a valid solution]. Therefore, we look for solution in the form
\begin{equation} \label{eq:Vz}
  V=i \pi/2 + z,
  \qquad
  |z|\ll 1.
\end{equation}
Equation \eqref{eq:V} immediately yields
\begin{equation} \label{eq:z}
  z = \frac{\pi}{2^{3/4} \sqrt{\delta_0}} \sqrt{\sqrt{1-\varepsilon} - i\sqrt{\varkappa}}.
\end{equation}
This solution is valid as long as $|z|\ll 1$, i.e., at
\begin{equation} \label{eq:cond_rigid}
  |1-\varepsilon|,\varkappa \ll \delta_0^2.
\end{equation}

The DoS \eqref{eq:DoSgen} at the outer surface of the N layer is then given by\footnote{In the vicinity of $\varepsilon=1$, quantity $(1-\varepsilon)^{1/2}$ is real (positive) at $\varepsilon<1$ and imaginary at $\varepsilon>1$. In the latter case, the branch of the complex function corresponding to the retarded Green functions that we work with, implies
\[
(1-\varepsilon)^{1/2} = e^{-i\pi/2} (\varepsilon-1)^{1/2}.
\]
This allows us to write the results at $\varepsilon$ both smaller and larger than $1$.}
\begin{equation}
  \nu_N(\varepsilon) = \frac{\pi}{\delta_0} \Re z
   = \frac{\pi^2 \varkappa^{1/4}}{2^{5/4} \delta_0^{3/2}} \times
  \left\{
   \begin{array}{ll}
   \sqrt{\sqrt{\frac{1-\varepsilon}{\varkappa}} + \sqrt{\frac{1-\varepsilon}{\varkappa} +1}}, & \varepsilon<1, \\
   \sqrt{\sqrt{\frac{\varepsilon-1}{\varkappa}} +1}, & \varepsilon>1.
   \end{array}
  \right.
\end{equation}

Two limiting cases can be further distinguished here:

(i) At $|1-\varepsilon|\gg \varkappa$, we obtain
  \begin{equation} \label{eq:nuN1strong}
  \nu_N(\varepsilon) =\frac{\pi^2}{2^{5/4} \delta_0^{3/2}} \times
  \left\{
   \begin{array}{ll}
   2^{1/2} (1-\varepsilon)^{1/4}, & \varepsilon<1, \\
   (\varepsilon-1)^{1/4}, & \varepsilon>1.
   \end{array}
  \right.
  \end{equation}
  In particular, this result is applicable at $\varepsilon$ arbitrarily close to $1$ in the limit of \emph{absolutely rigid} boundary condition, $\varkappa = 0$.
  This result agrees with the energy dependence in Eq.\ (15) from Ref.\ \cite{Levchenko2008} (which was written there without a numerical coefficient), and we call this the \emph{full check-mark} peculiarity, see Fig.~\ref{fig:peculiarities}(a). Note that the peculiarity is asymmetric, with its left ``wing'' being steeper than the right one due to the additional $2^{1/2}$ factor in Eq.\ \eqref{eq:nuN1strong}.

(ii) At $|1-\varepsilon|\ll \varkappa$, we obtain
  \begin{equation} \label{eq:nuN1}
  \nu_N(\varepsilon) =\frac{\pi^2}{2^{5/4} \delta_0} \left( \frac{\varkappa}{\delta_0^2} \right)^{1/4}
  \left( 1 + \frac{1}{2} \sqrt{\frac{|1-\varepsilon|}{\varkappa}} \right).
  \end{equation}
  So, at $\varepsilon=1$, we find finite DoS at any nonzero $\varkappa$. The peculiarity still has a form of a check mark, which we call the \emph{elevated check mark}, see Fig.~\ref{fig:peculiarities}(b). In addition to finite $\nu_N(1)$ value, it differs from the full check-mark peculiarity by how the DoS varies as $\varepsilon$ deviates from $1$: as $|1-\varepsilon|^{1/2}$ in the case of elevated check mark, in contrast to  $|1-\varepsilon|^{1/4}$ in the case of full check mark. In addition, the elevated check-mark peculiarity is symmetric.

The DoS value $\nu(1)$ following from Eq.\ \eqref{eq:nuN1} can be both small and large [since the small parameter $( \varkappa / \delta_0^2)^{1/4}$ is divided by the small parameter $\delta_0$].

\subsubsection{\texorpdfstring{$E_g$}{Eg} and \texorpdfstring{$E\to E_g$}{E->Eg}}

At $\varepsilon<\varepsilon_g$, Eq.\ \eqref{eq:V} has a real solution $V$ so that the DoS \eqref{eq:DoSgen} turns to zero.
At the same time, its left-hand side (l.h.s.) has a minimum for real $V$. As $\varepsilon$ rises above $\varepsilon_g$, the r.h.s\ of the equation falls below this minimum, and the solution $V$ becomes complex providing finite DoS.

In order to find the minimal value of the l.h.s.\ of Eq.\ \eqref{eq:V} at real $V$, we differentiate it and obtain equation
\begin{equation}
  \frac{\cosh V_0 - V_0 \sinh V_0}{V_0^3} = \sqrt{\frac{\varkappa}{2\delta_0^2}}
\end{equation}
for the position of the minimum.
At $\varkappa \ll \delta_0^2$, we can substitute the r.h.s.\ of this equation by zero, so the equation yields
\begin{equation}
  V_0 \tanh V_0 = 1
  \qquad \Rightarrow \qquad
  V_0 \approx 1.2.
\end{equation}
Substituting this into Eq.\ \eqref{eq:V}, we obtain
\begin{equation} \label{eq:Eg}
  \varepsilon_g = 1- \left( \frac{\cosh V_0}{V_0} \right)^4 \frac{\delta_0^2}{2} \approx 1-2.6 \delta_0^2.
\end{equation}
The applicability conditions \eqref{eq:Vcond} and \eqref{eq:epscloseto1} for this result are satisfied.
Equation \eqref{eq:Eg} reproduces the corresponding result from Ref.\ \cite{Levchenko2008} (see 
\ref{app:comparison} for additional comments on comparison with previous works).

We see that at $\varepsilon\to \varepsilon_g$ we can actually neglect the first term in Eq.\ \eqref{eq:V}, reducing the equation to
\begin{equation} \label{eq:EgVsimpl}
  \frac{\delta_0^{1/2} \cosh V}{2^{1/4} V} = (1-\varepsilon)^{1/4}.
\end{equation}
In order to find the DoS near the gap edge, we write the solution at $\varepsilon \to \varepsilon_g+0$ as $V=V_0+V_1$.
Expanding Eq.\ \eqref{eq:EgVsimpl} with respect to $V_1$, we find
\begin{equation}
 V_1 = i \left( V_0 / \cosh V_0 \right)^2 \sqrt{\varepsilon - \varepsilon_g} / \delta_0,
\end{equation}
and hence
\begin{equation} \label{eq:nuN1Eg}
\nu_N(\varepsilon) = (2V_0 / \delta_0) \Im V_1 = C_0 \sqrt{\varepsilon - \varepsilon_g} / \delta_0^2,
\end{equation}
where
\begin{equation}
C_0 = 2 V_0^3 / \cosh^2 V_0 \approx 1.05.
\end{equation}
This result is valid while $|V_1|\ll V_0$, i.e., at $(\varepsilon - \varepsilon_g) \ll \delta_0^2 \sim (1-\varepsilon_g)$.

Our result \eqref{eq:nuN1Eg} agrees with the energy dependence in Eq.\ (14) from Ref.\ \cite{Levchenko2008} (which was written there without a numerical coefficient).

\subsection{Limit of soft superconductor, \texorpdfstring{$\varkappa \gg \delta_0^2$}{kappa>>delta0squared}}

\subsubsection{\texorpdfstring{$E\to\Delta_0$}{E->Delta0}}

At large $\varkappa$ and $\varepsilon\to 1$, Eq.\ \eqref{eq:V} is solved by such $V$ that $|V|\ll 1$.
Equation \eqref{eq:V} is therefore simplified as
\begin{equation} \label{eq:Vweak}
\sqrt{\varkappa} V^3 - \sqrt{1-\varepsilon} V^2 + \delta_0/\sqrt{2} = 0.
\end{equation}

At $\varepsilon=1$, the solution is
\begin{equation}
  V_0= e^{i\pi/3} \left( \delta_0 / \sqrt{2\varkappa} \right)^{1/3},
\end{equation}
where we have chosen the root of $(-1)^{1/3}$ so, that the corresponding DoS is positive. Condition $|V|\ll 1$ is indeed satisfied under our current assumption $\varkappa \gg \delta_0^2$. At the same time, condition \eqref{eq:Vcond} requires that $\varkappa$ is not too large. The two conditions together have the form
\begin{equation} \label{eq:conditions}
\delta_0^2 \ll \varkappa \ll 1/\delta_0 .
\end{equation}
So, Eq.\ \eqref{eq:V} allows us to describe the crossover between the regimes of rigid and soft superconductor;
however, if the superconductor becomes too soft (too large $\varkappa$ such that $\varkappa\gtrsim 1/\delta_0$), then our solution becomes inapplicable [since condition \eqref{eq:Vcond} is violated, hence Eq.\ \eqref{eq:V} becomes inapplicable].

At $\varepsilon\to 1$, we write the solution with a small correction as
\begin{equation}
 V = V_0 + V_1,
\end{equation}
and Eq.\ \eqref{eq:Vweak} immediately yields
\begin{equation}
 V_1 = (1/3) \sqrt{(1-\varepsilon)/\varkappa}.
\end{equation}
This solution is valid if $|V_1|\ll |V_0|$, i.e., at
\begin{equation} \label{eq:cond_soft}
 |1-\varepsilon| \ll (\varkappa \delta_0)^{2/3} \sim \delta_0^2 \left( \varkappa / \delta_0^2 \right)^{2/3}.
\end{equation}

The DoS \eqref{eq:DoSgen} at the outer surface of the N layer is given by
\begin{equation} \label{eq:nuN1vert}
  \nu_N(\varepsilon) = \frac{\Im (V_0^2 + 2 V_0 V_1)}{\delta_0}
   = \frac{3^{1/2}}{2^{4/3}( \varkappa\delta_0)^{1/3}} + \frac{\sqrt{|1-\varepsilon|}}{3\cdot 2^{1/6}( \varkappa\delta_0)^{2/3}}  \times
  \left\{
   \begin{array}{ll}
   3^{1/2}, & \varepsilon<1, \\
   -1, & \varepsilon>1.
   \end{array}
  \right.
\end{equation}
This result corresponds to the \emph{vertical peculiarity} \cite{Fominov2019.PhysRevB.100.224513}, see Fig.~\ref{fig:peculiarities}(c). Note that $\nu_N(1)\gg 1$.
The peculiarity is asymmetric, with its left ``wing'' being steeper than the right one due to the additional $3^{1/2}$ factor in Eq.\ \eqref{eq:nuN1vert}.
At $\varkappa=1$, Eq.\ \eqref{eq:nuN1vert} reproduces Eq.\ (53) from Ref.\ \cite{Fominov2019.PhysRevB.100.224513} (see 
\ref{app:comparison} for comments on this comparison).

So, as $\varkappa$ grow and we cross over from the rigid to soft limit, the DoS given by Eq.\ \eqref{eq:nuN1} evolves into Eq.\ \eqref{eq:nuN1vert}.
Note that $\nu_N(1)$ as a function of $\varkappa$ turns out to be nonmonotonic: it grows at small $\varkappa$ in the rigid-S regime [according to Eq.\ \eqref{eq:nuN1}] but then decreases at larger $\varkappa$ in the soft-S regime [according to Eq.\ \eqref{eq:nuN1vert}]. This decrease can be viewed as resulting from softening of the superconductor (suppression of superconductivity in the S part near the interface). The maximal value $\nu_N(1) \sim 1/\delta_0$ is achieved at the crossover between the rigid and soft regimes, i.e., at $\varkappa\sim \delta_0^2$.

\subsubsection{\texorpdfstring{$E_g$}{Eg} and \texorpdfstring{$E\to E_g$}{E->Eg}}

Equation \eqref{eq:Vweak} can also be used for finding the gap $\varepsilon_g$ [assuming that it corresponds to $|V|\ll 1$].
To this end, we should look for disappearance of real solutions in Eq.\ \eqref{eq:Vweak}. The function in its l.h.s.\ starts from a positive value at $V=0$ and then decreases quadratically at very small positive $V$, finally starting to increase at larger $V$ due to the $V^3$ contribution. The position of the minimum is
\begin{equation}
  V_0 = (2/3) \sqrt{(1-\varepsilon)/\varkappa}.
\end{equation}
Requiring that the minimal value of the function is equal to $0$ (disappearance of real solutions), we obtain
\begin{equation} \label{eq:varepsilong}
  \varepsilon_g = 1- \frac{3^2 (\varkappa \delta_0)^{2/3}}{2^{5/3}} .
\end{equation}
The applicability conditions for this result [$|V|\ll 1$ and Eqs.\ \eqref{eq:epscloseto1} and \eqref{eq:Vcond}] are satisfied
due to Eq.\ \eqref{eq:conditions}.
At $\varkappa=1$, Eq.\ \eqref{eq:varepsilong} reproduces Eq.\ (32) from Ref.\ \cite{Fominov2019.PhysRevB.100.224513} (see 
\ref{app:comparison} for comments on this comparison).

Next, at $\varepsilon\to \varepsilon_g+0$, we write the solution as $V=V_0+V_1$. Expanding Eq.\ \eqref{eq:Vweak} with respect to $V_1$ and $(\varepsilon - \varepsilon_g)$, we find the relation between them:
\begin{equation}
 V_1 = i \sqrt{2(\varepsilon - \varepsilon_g)/9\varkappa},
\end{equation}
and hence
\begin{equation} \label{eq:nuNnearEg}
\nu_N(\varepsilon) = \frac{2V_0}{\delta_0} \Im V_1 = \frac{2^{5/3} \sqrt{\varepsilon - \varepsilon_g}}{3(\varkappa \delta_0)^{2/3}}.
\end{equation}
This result is valid at $|V_1|\ll V_0$, i.e., at $(\varepsilon - \varepsilon_g)\ll (1-\varepsilon_g)$.
It agrees with the energy dependence in Eq.\ (17a) from Ref.\ \cite{Levchenko2008} (which was written there without a numerical coefficient).
At $\varkappa=1$, Eq.\ \eqref{eq:nuNnearEg} reproduces Eq.\ (41) from Ref.\ \cite{Fominov2019.PhysRevB.100.224513} (see 
\ref{app:comparison} for comments on this comparison).

\section{Disappearance of the full check-mark peculiarity at \texorpdfstring{$\varkappa=0$}{kappa=0} with increasing \texorpdfstring{$d_N$}{dN}}
\label{sec:FiniteThickness}

In Section~\ref{sec:SNwithinvprox}, we have seen that while the full check-mark peculiarity of the DoS \cite{Levchenko2008} is realized at $\varkappa=0$ in the limit of thin N layer, it is ``unstable'' with respect to finite value of $\varkappa$ (immediately transforming to the elevated check-mark peculiarity). It is then natural to ask whether the full check-mark peculiarity at $\varkappa=0$ is stable with respect to increasing thickness $d_N$ of the N layer (or increasing $\delta_0$ which is a dimensionless thickness-dependent parameter). In this section, we address this question.

We will actually see that the full check-mark behavior [with $\nu_N(1)=0$] disappears only at $\delta_0 \sim 1$.
Therefore, the consideration of Section~\ref{sec:thinN} based on the assumption of thin N layer is not applicable for our current purpose since condition \eqref{eq:delta0<<1} is violated.
We thus have to consider the set of exact equations \eqref{eq:psiNUs}--\eqref{eq:psiNbc}.
At the same time, at $\varkappa=0$, the Green function in the S part coincides with its bulk value, hence $\psi_S(X)=0$,
and the boundary conditions \eqref{eq:psi_N(0)} and \eqref{eq:psi_N'(0)} are substituted by a single requirement
\begin{equation} \label{eq:psi_N(0)_rigid}
\psi_N(0) = \ln \sqrt{(1+\varepsilon)/(1-\varepsilon)}
\end{equation}
(we have expressed $\arctanh\varepsilon$ in the logarithmic form).
So, we have to consider the set of equations \eqref{eq:psiNUs}, \eqref{eq:psiNbc}, and \eqref{eq:psi_N(0)_rigid}.

The Usadel equation \eqref{eq:psiNUs} has the first integral
\begin{equation}
(\psi_N')^2 /4 + \varepsilon \delta_0 \sinh \psi_N = \varepsilon \delta_0 \sinh \psi_N(1),
\end{equation}
which yields
\begin{equation}
2\sqrt{ \varepsilon\delta_0} X = \int_{\psi_N(0)}^{\psi_N(X)} \frac{d\psi}{\sqrt{\sinh\psi_N(1) - \sinh\psi}}.
\end{equation}

Introducing a new function
\begin{equation}
w(X) = e^{-\psi_N(X)},
\end{equation}
we obtain
\begin{equation} \label{eq:1intw}
\sqrt{2\varepsilon\delta_0 } X = -\int_{w(0)}^{w(X)} \frac{dw}{w \sqrt{\left[ w^{-1}(1) - w(1) \right] - \left[ w^{-1} - w \right]}},
\end{equation}
with
\begin{equation} \label{eq:w(0)}
w(0) =\sqrt{(1-\varepsilon) / (1+\varepsilon)}.
\end{equation}
The DoS \eqref{eq:nu_N} in new notations is
\begin{equation} \label{eq:DoSw}
\nu_N(\varepsilon) = \Im \left[ w^{-1}(1) - w(1) \right]/2.
\end{equation}

To make connection with our previous results of Section~\ref{sec:SNwithinvprox}, we note that the limit of thin N layer corresponds to $|w|\ll 1$ [according to Eq.\ \eqref{eq:approx}].
Neglecting $w$ in comparison to $w^{-1}$ under the integral in Eq.\ \eqref{eq:1intw}, we can explicitly perform the integration.
Considering $X=1$, we explicitly find $w(1)$, and then obtain $w(X)$ which is, of course, equivalent to Eq.\ \eqref{eq:psiN5}.
At $\varepsilon=1$, the result has the form
\begin{equation} \label{eq:w(x)}
w(X) = - (2\delta_0/ \pi^2) \sin^2 \left( \pi X / 2 \right).
\end{equation}
Note that this function is real and negative. According to Eq.\ \eqref{eq:DoSw}, real $w$ yields zero DoS.

Now, we return to the full equation \eqref{eq:1intw}, which we write at $X=1$ as
\begin{equation} \label{eq:f(w,w1)}
\sqrt{2 \varepsilon\delta_0} = -\int_{w(0)}^{w(1)}
dw \sqrt{f (w;w(1))},
\end{equation}
with
\begin{equation} \label{eq:f}
f(w;w(1)) = \frac{1}{w^3 -w + w^2 \left[ w^{-1}(1) - w(1) \right]}.
\end{equation}
We want to consider the possibility to find real solution $w(1)$ at $\varepsilon=1$ as the thickness of the N layer increases (i.e., as $\delta_0$ increases), and this solution should be a result of integration along the real axis of $w$ (otherwise, the DoS cannot be zero). The form of $f(w;w(1))$ [where $w(1)$ is a parameter] is essential in this respect.
Since $w(0)=0$ at $\varepsilon=1$, the positive sign of the r.h.s.\ in Eq.\ \eqref{eq:f(w,w1)} should be ensured by $w(1)<0$ [note that this statement agrees with the explicit thin-layer result \eqref{eq:w(x)}].

\begin{figure}[t]
    \centerline{\includegraphics[width=0.6\columnwidth]{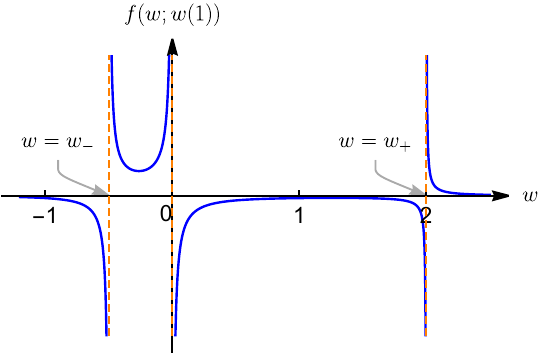}}
    \caption{Typical form of $f(w; w(1))$ defined by Eq.\ \eqref{eq:f} [this specific plot corresponds to $w(1) = -0.5$].
    As a function of $w$, the $f(w;w(1))$ function has three poles: one at $w=0$ and two at $w_{\pm}$ of opposite signs.
    At negative $w(1)$, we have $w_- = w(1)$ and $w_+ = -w^{-1}(1)$. Between $w_-$ and $0$, the function always has a positive-valued U shape, while between $0$ and $w_+$ it has the inverted (negative-valued) U shape. At $w<w_-$ and $w>w_+$, the function is monotonic.}
  \label{fig:fwDdep}
\end{figure}

A typical form of $f(w;w(1))$ is presented in Fig.~\ref{fig:fwDdep}.
With this information, we can understand how Eq.\ \eqref{eq:f(w,w1)} works when we are looking for its real solution $w(1)$.

At $\varepsilon>1$, the real-valued solution of Eq.\ \eqref{eq:f(w,w1)} is impossible since $w(0)$ is imaginary, see Eq.\ \eqref{eq:w(0)}.
At $\varepsilon<1$, the real-valued solution is impossible due to a more subtle reason. In this case, $w(0)>0$, hence integration along the positive interval of the $w$ axis in Eq.\ \eqref{eq:f(w,w1)} inevitably yields an imaginary contribution since $f(w;w(1))<0$ in this region, see Fig.~\ref{fig:fwDdep}.

\begin{figure}[t]
    \centerline{\includegraphics[width=0.6\columnwidth]{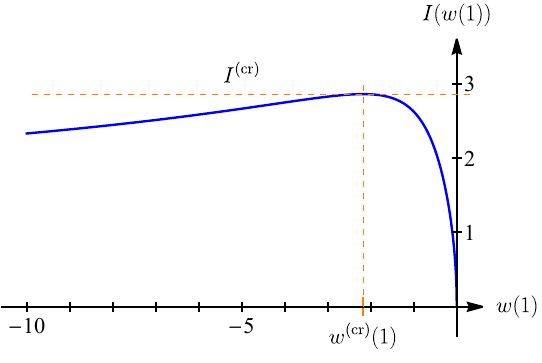}}
    \caption{$I(w(1))$ function from Eq.\ \eqref{eq:IwDef}.}
  \label{fig:Iw}
\end{figure}

At $\varepsilon=1$, we have $w(0)=0$, and the real-valued solution of Eq.\ \eqref{eq:f(w,w1)} is possible since the integration is now from $0$ to a negative $w(1)$.
Equation \eqref{eq:f(w,w1)} simplifies as
\begin{equation} \label{eq:IwDef}
\sqrt{2\delta_0} = I(w(1)) \equiv  \int_{w(1)}^{0} dw \sqrt{f(w;w(1))},
\end{equation}
and we maximize the integral $I$ as a function of negative $w(1)$. The form of $I(w(1))$ is presented in Fig.~\ref{fig:Iw}. The maximal value $I^\mathrm{(cr)} \approx 2.86$
is achieved at $w^\mathrm{(cr)}(1)\approx -2.18$.
This implies the maximal value of $\delta_0$ [in the l.h.s.\ of Eq.\ \eqref{eq:IwDef}] and the corresponding critical value $d_N^\mathrm{(cr)}$ such that $\nu_N(1)$ becomes finite at $d>d_N^\mathrm{(cr)}$:
\begin{equation} \label{eq:dNcr}
d_N^\mathrm{(cr)} / \xi_N \approx 2.86.
\end{equation}

Equation \eqref{eq:IwDef} can also be used in order to calculate how $\nu_N(1)$ depends on $d_N$ just above the critical value $d^\mathrm{(cr)}_N$.
For that, we expand $I(w(1))$ near its maximum,
\begin{equation}
	I(w(1)) = I^\mathrm{(cr)} + \left(  I'' / 2 \right) \left[ w(1) -  w^\mathrm{(cr)}(1) \right]^2,
\end{equation}
where
$I'' \equiv I''_{ww}(w^\mathrm{(cr)}(1)) \approx -0.15$ is found numerically.
Expressing $\delta_0$ in terms of $d_N$ [see Eq.\ \eqref{eq:ETh_delta0}], we arrive at
\begin{equation}
	w(1) = w^\mathrm{(cr)}(1) - i \sqrt{(2 / | I''|) \left[ d_N -d^\mathrm{(cr)}_N \right] /\xi_N},
\end{equation}
where the minus sign in front of the square root is chosen in order to obtain the positive DoS \eqref{eq:DoSw}:
\begin{equation} \label{eq:RigidDdep}
	\nu_N(1)
    =\left\{
   \begin{array}{ll}
		0, & d_N < d^\mathrm{(cr)}_N , \\
		 2.21 \sqrt{\left[ d_N -d^\mathrm{(cr)}_N \right]/ \xi_N}, & d_N > d^\mathrm{(cr)}_N.
	\end{array}
    \right.
\end{equation}

\section{Numerical results for the DoS}
\label{sec:numerics}

Numerically, we solve Eqs.\ \eqref{eq:psiNUs}--\eqref{eq:psiNbc} and then calculate the DoS given by Eq.\ \eqref{eq:nu_N}.
This calculation has wider region of applicability than our analytical calculations in Section~\ref{sec:SNwithinvprox} since the numerical procedure is not limited by the condition of thin N layer [$d_N\ll \xi_N$ or Eq.~\eqref{eq:delta0<<1}].

We use a numerical procedure based on the Python solver scipy.integrate.solve\_bvp from the SciPy library \cite{Virtanen2020}.

\subsection{Absolutely rigid limit}

\begin{figure}
\begin{center}
    \includegraphics[width=0.7\columnwidth]{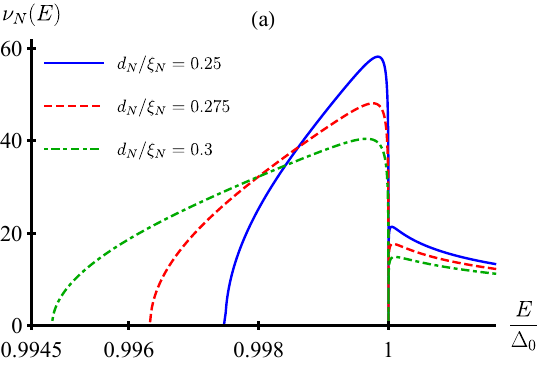}\\
    \bigskip
	\includegraphics[width=0.7\columnwidth]{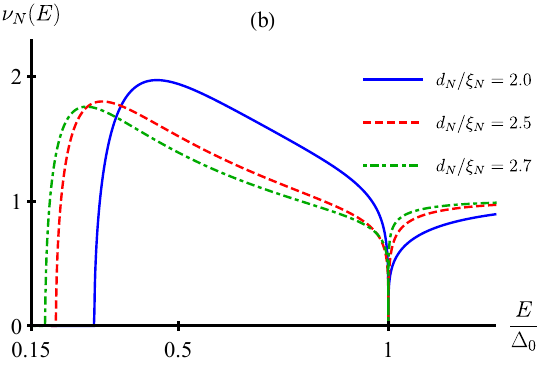}
\end{center}
\caption{Numerical results for the surface DoS in the absolutely rigid limit. Different curves correspond to different thicknesses $d_N$ of the N layer. All the curves demonstrate the gap $E_g<\Delta_0$ and the full check-mark peculiarity with $\nu_N=0$ at $E=\Delta_0$ \cite{Levchenko2008}. (a)~Thin-layer limit corresponding to the analytical treatment of Section~\ref{sec:rigid}. (b)~Regime of moderate $d_N/\xi_N$. The full check-mark peculiarity becomes wider.}
  \label{fig:DOSRigid}
\end{figure}

Although the limit of absolutely rigid boundary conditions (or absolutely rigid superconductor) defined by condition $\varkappa = 0$ can hardly be achieved experimentally, it is often assumed in theoretical calculations due to its simplicity. At the same time, as we have seen above, this limit is special from the point of view of the DoS behavior since it allows achieving the full check-mark peculiarity.

As we have discussed in Section~\ref{sec:FiniteThickness}, in the limit $\varkappa= 0$,
the exact equations of our theory reduce to Eqs.\ \eqref{eq:psiNUs}, \eqref{eq:psiNbc}, and \eqref{eq:psi_N(0)_rigid}.
This boundary value problem by can be numerically solved \cite{Virtanen2020} in a straightforward manner without any additional assumptions about the thickness of the N layer.
The numerical results are shown in Fig.~\ref{fig:DOSRigid}. We indeed see the full check-mark behavior with the property $\nu_N = 0$ at $E = \Delta_0$, in accordance with Eq.\ \eqref{eq:nuN1strong}.

The width $\delta\varepsilon$ of the check-mark peculiarity is rather small at small $\delta_0$ (i.e., at small $d_N/\xi_N$). According to Eq.\ \eqref{eq:cond_rigid}, the width should be $\delta\varepsilon \sim \delta_0^2$, i.e., of the same order as the gap suppression $(1-\varepsilon_g)$, see Eq.\ \eqref{eq:Eg}. The maximum of the DoS reached at some energy $\varepsilon_m$ between $\varepsilon_g$ and $1$ is of the order of $1/\delta_0 \gg 1$, while at energies of the order of $(1+\delta\varepsilon)$, the DoS crosses over to the BCS behavior $\nu_\mathrm{BCS} =\varepsilon/\sqrt{\varepsilon^2-1}$ \cite{Levchenko2008}. Numerically, in Fig.~\ref{fig:DOSRigid}(a), we see that $\varepsilon_m$ is much closer to $1$ than to $\varepsilon_g$. This additional narrowing of the check-mark peculiarity is due to the fact that it is actually not just the $|1-\varepsilon|/\delta_0^2$ parameter that must be small, but also the power $1/4$ of this parameter, see Eqs.\ \eqref{eq:Vz} and \eqref{eq:z}.

\begin{figure}[t]
    \centerline{\includegraphics[width=0.7\columnwidth]{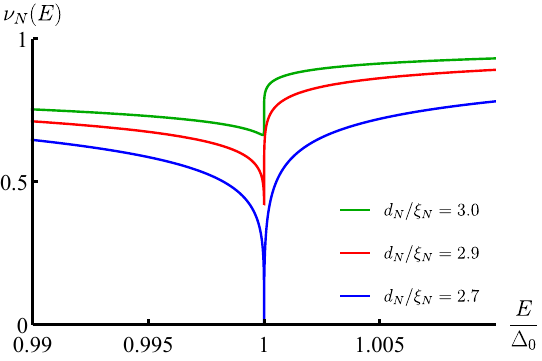}}
    \caption{Elevation of the check-mark peculiarity at thicknesses above the critical one. In accordance with Eq. \eqref{eq:dNcr}, the blue curve demonstrates the full check-mark behavior ($d_N< d^\mathrm{(cr)}_N$), while the red and green curves demonstrate the elevated check marks ($d_N> d^\mathrm{(cr)}_N$).}
    \label{fig:DoSNdep}
\end{figure}

At moderate values of $d_N/\xi_N$, the full check-mark peculiarity becomes wider, see Fig.~\ref{fig:DOSRigid}(b). Upon further increasing of $d_N$, above the critical value given by Eq. \eqref{eq:dNcr}, the check mark detaches from zero, and we obtain an elevated check-mark behavior, see Fig.~\ref{fig:DoSNdep}.

\subsection{Limits of rigid and soft superconductor}
\label{sec:RigidSoft}

\begin{figure}
\begin{center}
    \includegraphics[width=0.55\columnwidth]{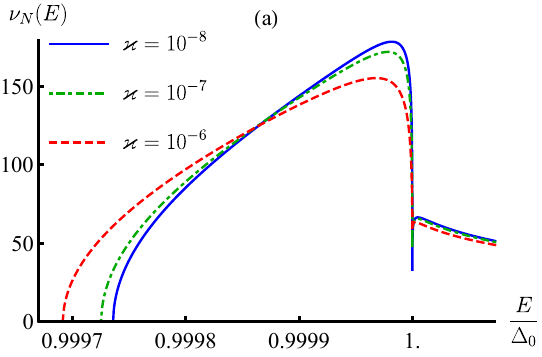}\\
    \bigskip
    \includegraphics[width=0.55\columnwidth]{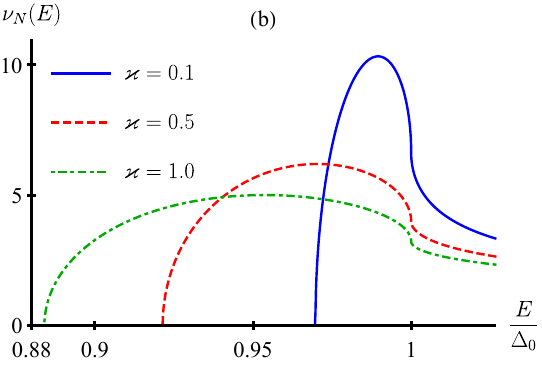}\\
    \bigskip
    \includegraphics[width=0.55\columnwidth]{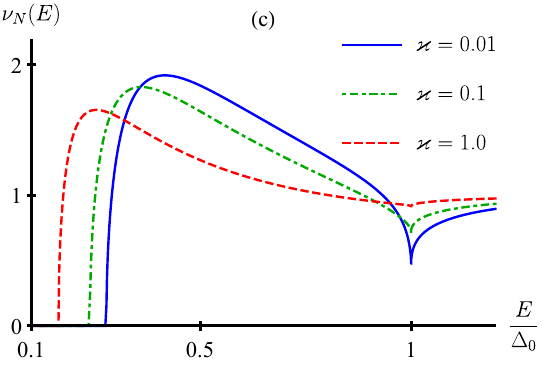}
\end{center}
    \caption{Surface DoS at nonzero $\varkappa$. Panels~(a) and~(b) correspond to the limit of thin N layer ($\delta_0 = 10^{-2}$).
    (a)~Rigid-S regime: narrow elevated check-mark peculiarities.
    (b)~Soft-S regime: vertical peculiarities.
    Panel~(c) corresponds to moderate thickness of the N layer ($\delta_0 = 2$). In this case, the check-mark peculiarities are much wider than in the thin-N limit.}
   \label{fig:DoSWeakStrong}
\end{figure}

At nonzero $\varkappa$, the check-mark peculiarity is immediately elevated, see Fig.~\ref{fig:DoSWeakStrong}(a). This happens at really small values of $\varkappa$ which are still much smaller than $\delta_0^2$, in accordance with Section~\ref{sec:rigid}. This implies that the \emph{full} check-mark behavior is hardy achievable in realistic structures of the considered type.

Moreover, at $\varkappa \sim \delta_0^2$, the check-mark behavior disappears completely turning into the vertical peculiarity, see Fig.~\ref{fig:DoSWeakStrong}(b). Therefore, \emph{any} check-mark behavior is hardly achievable in realistic structures of the considered type in the limit on thin N layer (since it requires very small $\varkappa$ values which are problematic for experimental realization).

The width of the vertical peculiarity is visibly larger than the width of the check-mark peculiarity (at the same value of $\delta_0$). This can be understood from our analytical consideration since the energy width of the vertical peculiarity is equal to the check-mark width $\delta_0^2$ multiplied by a large parameter $\left( \varkappa / \delta_0^2 \right)^{2/3}$, see Eq.\ \eqref{eq:cond_soft}.

Although the check marks shown in Fig.~\ref{fig:DoSWeakStrong}(a) are very narrow, this qualitative behavior is realized also for larger thicknesses $d_N$ which are beyond the analytical treatment of Section~\ref{sec:SNwithinvprox}. Numerically, we see that at moderate $d_N/\xi_N$, the check marks are much wider and clearly visible, see Fig.~\ref{fig:DoSWeakStrong}(c).

\subsection{Switching between the check-mark and vertical peculiarity}
\label{sec:SwitchingPecCheck}

\begin{figure}
\begin{center}
\includegraphics[width=0.55\columnwidth]{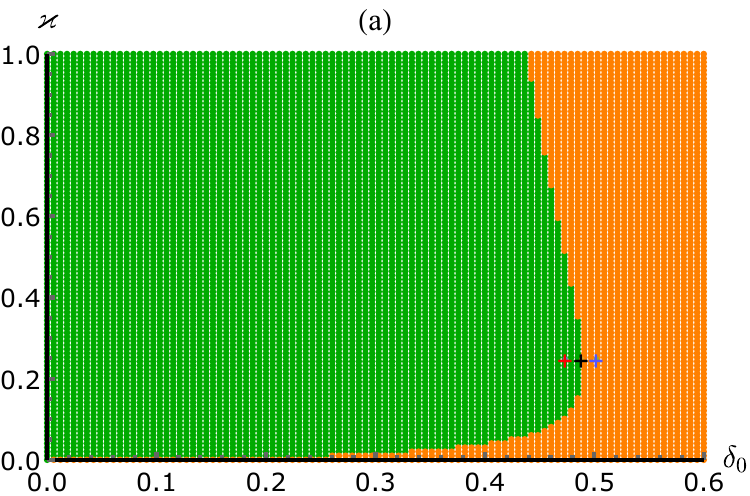}\\
    \bigskip
    \includegraphics[width=0.55\columnwidth]{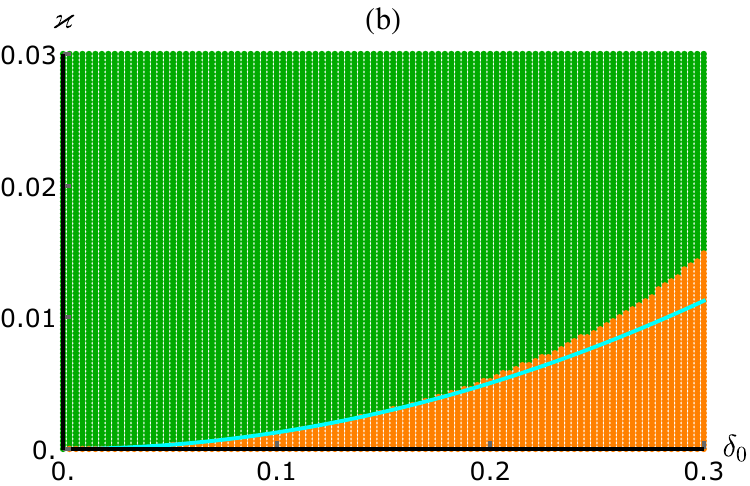}\\
    \bigskip
    \includegraphics[width=0.55\columnwidth]{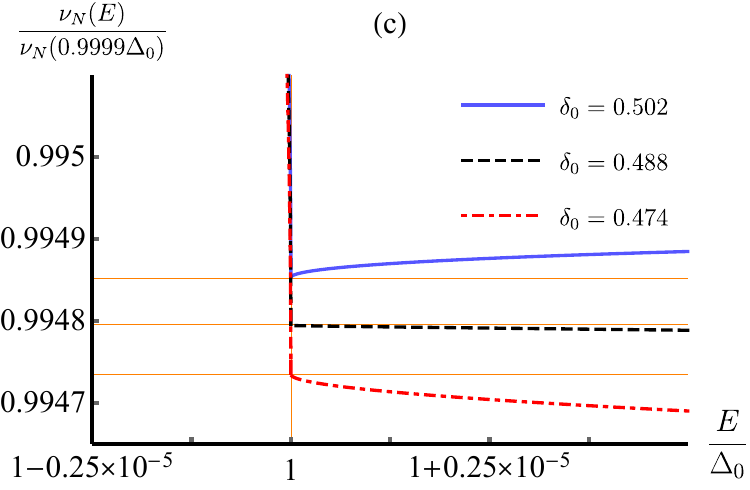}
\end{center}
    \caption{(a)~Phase diagram for the type of peculiarity of the DoS at $\varepsilon=1$. The orange region corresponds to the check-mark peculiarity. The green region corresponds to the vertical peculiarity.
    (b)~Zoomed region of small $\delta_0$ and $\varkappa$. The boundary between the orange and green regions is approximately described by the parabolic dependence $\varkappa=\delta_0^2/8$ (cyan line).
    The three crosses (red, black, and blue) in panel~(a) correspond to $\varkappa=0.25$ and three different values of $\delta_0$ ($0.474$, $0.488$, and $0.502$, respectively). The form of the DoS (in a very close vicinity of $\varepsilon=1$) corresponding to the three crosses is shown in panel (c).
    Note the change of sign of the $d\nu_N/d\varepsilon$ derivative at $\varepsilon>1$; this sign distinguishes the two regions of the phase diagram in panels~(a) and~(b).}
    \label{fig:PhaseDiagram1}
\end{figure}

How does the switching between the check-mark behavior [rigid superconductor, see Eq.\ \eqref{eq:nuN1}] and the vertical peculiarity [soft superconductor, see Eq.\ \eqref{eq:nuN1vert}] occur? As $\varkappa$ grows, the DoS $\nu_N(\varepsilon)$ always has a negative derivative at $\varepsilon=1-0$, while positive values of the derivative at $\varepsilon=1+0$ change to negative ones, see Fig.~\ref{fig:PhaseDiagram1}.
The analysis of Section~\ref{sec:SNwithinvprox} demonstrates that in the regime of small $\delta_0$, the crossover takes place at $\varkappa \sim \delta_0^2$.

At the same time, we can try to naively approximate the crossover point equating the two results for the DoS, given by Eqs.\ \eqref{eq:nuN1} and \eqref{eq:nuN1vert}, at $\varepsilon=1$. This yields
$\delta_0^2 / \varkappa = \left( 2\pi^{24} / 3^6 \right)^{1/7} \approx 22$.
Numerical results of Figs.~\ref{fig:PhaseDiagram1}(a) and~\ref{fig:PhaseDiagram1}(b) demonstrate that
the actual value is about three times smaller: the crossover line is well described by the parabolic dependence $\varkappa = \delta_0^2/8$ at $\delta_0 < 0.2$.

Interestingly, Fig.~\ref{fig:PhaseDiagram1}(a) demonstrates that the green region can be limited from above. A hint to such a possibility is contained in the applicability conditions \eqref{eq:conditions} for the vertical-peculiarity behavior.
Note that our analytical results for this type of behavior (corresponding to the green region in the figure) are valid only at $\varkappa\ll 1/\delta_0$.
At the same time, we do not consider larger $\varkappa$ within our analytical approach because self-consistency for the order parameter (which we have neglected) can become important in this limit.

The check-mark region (orange) becomes wider as $\delta_0$ grows, see Fig.~\ref{fig:PhaseDiagram1}(b). Moreover, at $\delta_0\approx 0.489$, the green region in Fig.~\ref{fig:PhaseDiagram1}(a) disappears completely.
This is already beyond applicability of our analytical theory (based on the assumption $\delta_0\ll 1$).
So, growth of $\delta_0$ stabilizes the (elevated) check-mark behavior.

\section{Discussion}
\label{sec:Discussion}

In this section, we discuss possible experimental implementations of the obtained results.

We have demonstrated that the type of the DoS peculiarity at $E=\Delta_0$ is determined by the relation between two dimensionless parameters, $\varkappa$ and $\delta_0$.
The materials-matching parameter $\varkappa$ [see Eq.\ \eqref{eq:varkappa}] does not depend on temperature $T$. At the same time, $\delta_0$ [see Eq.\ \eqref{eq:ETh_delta0}] contains $\Delta_0(T)$ and thus varies with temperature. This opens up experimental possibility to observe switching between qualitatively different types of the DoS behavior due to changing $T$. In terms of Figs.~\ref{fig:PhaseDiagram1}(a) and~\ref{fig:PhaseDiagram1}(b), the observation point moves horizontally to the right as temperature is lowered, so the vertical peculiarity can cross over to the check-mark peculiarity.

Experimental observation of the DoS peculiarities in SN junctions is a challenging task requiring a technique with high energy resolution.
At the same time, inelastic scattering or special types of pair-breaking disorder, which can effectively be described by the Dynes parameter \cite{Dynes1978.PhysRevLett.41.1509,Herman2016.PhysRevB.94.144508}, may wash out the peculiarities and hinder their observation.

Nevertheless, very-low temperature STM experiment by le~Sueur et al.\ \cite{leSueur2008.PhysRevLett.100.197002} contains signatures of the possible elevated check mark peculiarity at $E=\Delta_0$ (as evidenced by some of experimental curves in Figs.~2 and~3).
Experiment by Meschke et al.\ \cite{Meschke2011.PhysRevB.84.214514} demonstrated very high energy resolution by implementing tunneling spectroscopy of SN junctions with the help of a superconducting probe. The results evidenced a sharp drop of the DoS in the N part at $E=\Delta_0$ \cite{Meschke2011.PhysRevB.84.214514}.
In the context of the STM technique, superconducting probe (superconducting STM tip) significantly increases energy resolution \cite{Rodrigo2004}, which is advantageous for observing the DoS peculiarities.
We hope that our results will stimulate further experimental research in this direction.

\section{Conclusions}
\label{sec:Conclusions}

We have studied the surface DoS $\nu_N(E)$ in a diffusive SN system with half-infinite superconductor and transparent SN interface.
The strength of the proximity effect (both direct and inverse) is controlled by the materials-matching parameter $\varkappa$, see Eq.\ \eqref{eq:varkappa}.

In the limit of thin N layer, $d_N\ll \xi_N$, we have found three different types of the DoS peculiarity at $E=\Delta_0$.
(i)~At $\varkappa=0$ (absolutely rigid S, no inverse proximity effect), the peculiarity has the (full) check-mark form with $\nu(\Delta_0)=0$, see Eq.\ \eqref{eq:nuN1strong}. This form was predicted earlier in Ref.\ \cite{Levchenko2008}.
(ii)~At $\varkappa>0$ (rigid S, very weak inverse proximity effect), the check-mark is immediately elevated so that $\nu(\Delta_0)>0$, see Eq.\ \eqref{eq:nuN1}.
(iii)~At $\varkappa \gtrsim (d_N/\xi_N)^4$ (crossover to the soft S regime with essential inverse proximity effect), $\nu(E)$ gradually evolves to the vertical peculiarity, see Eq.\ \eqref{eq:nuN1vert}. This type of peculiarity was earlier obtained in Ref.\ \cite{Fominov2019.PhysRevB.100.224513}
(although in a different system which can only correspond to the $\varkappa=1$ case of our current theory, see 
\ref{app:comparison} for detail).
Regimes~(ii) and~(iii) correct earlier predictions of Ref.\ \cite{Levchenko2008}.

The elevated check-mark regime [regime~(ii)] has not been analytically described before, to the best of our knowledge. It is a ``missing element'' that describes continuous evolution of the DoS peculiarity with varying the $\varkappa$ parameter.

In the absolutely rigid limit ($\varkappa=0$), the full check-mark peculiarity is realized as long as $d_N$ is smaller than the critical thickness of the order of $\xi_N$, see Eq.\ \eqref{eq:dNcr}. At larger $d_N$, the check-mark is elevated.

In the above regimes (i)--(iii), we also calculate the energy gap $E_g$ and $\nu(E)$ in the vicinity of $E_g$. At small thickness $d_N$, the gap is only slightly smaller than $\Delta_0$.

Our results demonstrate that behavior of the DoS near $E=\Delta_0$ is very sensitive to the boundary conditions, and the full check-mark behavior can be easily destroyed (e.g., by finite $\varkappa$). At the same time, it has been recently shown that the SN interface in the form of constriction (quantum point contact) can stabilize this type of peculiarity stretching it into a secondary gap (``smile'' gap) in the DoS
\cite{Reutlinger2014.PhysRevLett.112.067001,Reutlinger2014.PhysRevB.90.014521,Yokoyama2017.PhysRevB.95.045411,Whisler2018.PhysRevB.97.224515}.
These results were obtained in setups with the N part represented by a chaotic cavity (quantum dot), implying the Green functions constant in space (0D limit).
It would be interesting to apply our 1D approach (taking into account spatial gradients) to the systems which are expected to demonstrate the secondary gap.
Another interesting open question is influence of non-ideal interface transmission on the peculiarity of the DoS in our planar-interface geometry.

\section*{Acknowledgements}
We thank A.\ Levchenko for useful discussions.
Ya.V.F.\ was supported by the Basic research program of HSE.
This work was also supported by the Foundation for the Advancement of Theoretical Physics and Mathematics ``BASIS''.

\appendix

\section{Applicability check}
\label{app:checks}

In the main text, we have checked that condition \eqref{eq:Vcond} was indeed satisfied for our solutions. This actually corresponds to condition \eqref{eq:approx} at $X=1$ written in different notations. At the same time, we still need to check validity of condition \eqref{eq:approx} \emph{everywhere} inside the N layer (not only at $X=1$). Below we demonstrate that it is indeed satisfied under our assumptions.

To this end, we rewrite Eq.\ \eqref{eq:psiN5} as
\begin{equation}
e^{\psi_N(X)} = e^{\psi_N(1)}/ \cosh^2 [ V(X-1)].
\end{equation}
Our solutions in Sections~\ref{sec:SNwithinvprox} and~\ref{sec:FiniteThickness} correspond to either $|V|\ll 1$ or $|V|\sim 1$, while $|e^{\psi_N(1)}| \gg 1$.
This guarantees that $|e^{\psi_N(X)}| \gg 1$ at any $X$ between $0$ and $1$.

\section{Comparison with previous works}
\label{app:comparison}

Here we present some details of comparison between the results of this paper and the results presented earlier in Refs.\ \cite{Levchenko2008} and \cite{Fominov2019.PhysRevB.100.224513}.

We actually solve the same problem as considered previously by Levchenko \cite{Levchenko2008} but obtain essentially different results. In our opinion, the discrepancy is a consequence of an uncontrolled approximation implicitly used in Ref.\ \cite{Levchenko2008}. Repeating the derivation, we notice that Eq.\ (9) in Ref.\ \cite{Levchenko2008} should additionally contain a term of the order of $\gamma/ u_S$  (in notations of Ref.\ \cite{Levchenko2008}) in its l.h.s. This term cannot generally be neglected, therefore the results of  Ref.\ \cite{Levchenko2008} should be reconsidered, and we do it in this paper.

In our notations, we can move the last term from the l.h.s.\ in Eq.\ \eqref{eq:V} to its r.h.s.\ and then square the equation obtaining
\begin{equation} \label{eq:inapp}
\varkappa \sinh^2 V = (1-\varepsilon) - \sqrt{2(1-\varepsilon)} \frac{\delta_0 \cosh^2 V}{V^2}
+ \frac{1}{2} \left( \frac{\delta_0 \cosh^2 V}{V^2} \right)^2.
\end{equation}
The derivation of Ref.\ \cite{Levchenko2008} effectively disregards the last term in the r.h.s.\ of Eq.\ \eqref{eq:inapp}.

Some resulting agreements and disagreements between our current work and Ref.\ \cite{Levchenko2008} have already been mentioned in the main text. Now we comment upon two less obvious comparisons:

(i)~Our result \eqref{eq:Eg} coincides with Eq.\ (13) from Ref.\ \cite{Levchenko2008} if we take into account that (a)~the Thouless energy $\epsilon_{Th}$ in Ref.\ \cite{Levchenko2008} is 4 times smaller than our $E_\mathrm{Th}$, and (b)~the numerical value $\mathcal{F}_m\approx 0.5$ in Ref.\ \cite{Levchenko2008} should be more accurately written as $\mathcal{F}_m \approx 0.469$ [so that the coefficient in front of $(\Delta_0/\epsilon_{Th})^2$ in Eq.\ (13) of Ref.\ \cite{Levchenko2008} is actually $1/128 \mathcal{F}_m^4 \approx 0.162$].

(ii)~Our result \eqref{eq:varepsilong} corrects Eq.\ (16) from Ref.\ \cite{Levchenko2008}. The result in Ref.\ \cite{Levchenko2008} was parametrically correct but with a wrong numerical coefficient in the r.h.s. [$2^{1/3}$ in our notations instead of $3^2/2^{5/3}$ as stated by our Eq.\ \eqref{eq:varepsilong}].

Finally, we comment upon comparison between our current results and results of our previous work \cite{Fominov2019.PhysRevB.100.224513}.
In that work, we studied peculiarity of the surface DoS at $E=\Delta_0$ in a different system, a superconductor with surface suppression of the BCS pairing constant $\lambda(x)$, see Fig.~\ref{fig:SN}.
This corresponds to \emph{weak} suppression of $\lambda(x)$ on a short spatial scale $r_c \ll \xi_S$ near the surface.
At the same time, an SN system with $d_N \ll \xi_N$ can also be considered as a system with short-scale surface suppression of $\lambda(x)$ but in the limit of \emph{strong} suppression.
As we explain in the main part of the paper, the most important physical assumption is the assumption of small spatial scale of the suppression.
This is exactly why comparison between our current results and the results of Ref.\ \cite{Fominov2019.PhysRevB.100.224513} is possible but only at a specific value $\varkappa=1$ (which corresponds to the system of Ref.\ \cite{Fominov2019.PhysRevB.100.224513} by definition).
In order to implement this comparison, we need to express the $d_1$ parameter from Ref.\ \cite{Fominov2019.PhysRevB.100.224513} in our current notations: $d_1 = d_N/\xi_N = \sqrt{2\delta_0}$.
As a result, we find that Eqs.\ \eqref{eq:nuN1vert}, \eqref{eq:varepsilong}, and \eqref{eq:nuNnearEg} at arbitrary values of $\varkappa$ reproduce Eqs.\ (53), (32), and (41) from Ref.\ \cite{Fominov2019.PhysRevB.100.224513}, respectively, if one identifies $d_1 = \sqrt{2 \varkappa \delta_0}$.

Note that our current result \eqref{eq:Eg} is parametrically different from the result (32) for the gap from our previous work \cite{Fominov2019.PhysRevB.100.224513}.
This is not surprising since the results of Ref.\ \cite{Fominov2019.PhysRevB.100.224513} correspond to $\varkappa=1$, i.e., to the limit of soft superconductor in our current terminology, while our current Eq.\ \eqref{eq:Eg} is obtained in the opposite limit of rigid superconductor.
Still, it is instructive to understand what exactly breaks down in the solution of Ref.\ \cite{Fominov2019.PhysRevB.100.224513} in this limit.
The point is that in the limit of rigid boundary conditions, the spatial scale for the Green function's variation in the entire system is not $\xi_E \gg d_N$ [see Eq.\ \eqref{eq:xiE}] anymore but becomes equal to $d_N$ itself (since the solution is rigidly fixed in the S part). The effective delta-functional boundary condition for $\psi$, employed in Ref.\ \cite{Fominov2019.PhysRevB.100.224513}, is not applicable in this situation.

\bibliographystyle{elsarticle-num}

\end{document}